\documentclass{emulateapj}
\usepackage{graphicx}
\usepackage{amssymb}
\usepackage{natbib}
\bibpunct{(}{)}{;}{a}{}{,}

\shorttitle{VLA imaging of isotopic SiO maser emission in Orion BN/KL}
\shortauthors{Goddi et al. 2009}

\newcommand{\tr}{$^{30}$SiO}
\newcommand{\tn}{$^{29}$SiO}
\newcommand{\te}{$^{28}$SiO}

\newcommand{\kms}{km~s$^{-1}\,$}

\newcommand{\pas}{$\rlap{.}^{\prime\prime}$}

\begin{document}

\title{Maser emission from SiO isotopologues traces the innermost 100~AU around Radio Source~I in Orion BN/KL}

\author{C. Goddi, L. J. Greenhill}
\affil{Harvard-Smithsonian Center for Astrophysics,
    60 Garden Street, Cambridge, MA 02138}

\and

\author{C. J. Chandler}
\affil{National Radio Astronomy Observatory, P.O. Box O, Socorro, NM 87801}

\and

\author{E. M. L. Humphreys, L. D. Matthews\footnote{Present address:
MIT Haystack Observatory, Westford, MA 01886}}
\affil{Harvard-Smithsonian Center for Astrophysics,
60 Garden Street, Cambridge, MA 02138}

\and

\author{M.D. Gray}
\affil{Jodrell Bank Centre for Astrophysics, Alan Turing Building, School of Physics and Astronomy, University of Manchester, Manchester M13 9PL}

\begin{abstract}
We have used the Very Large Array (VLA) at 7~mm wavelength to image five rotational transitions  ($J=1-0$) from three SiO isotopologues towards Orion BN/KL: 
$^{28}$SiO $v=0,1,2$; and $^{29}$SiO and $^{30}$SiO $v=0$.
For the first time, we have mapped the $^{29}$SiO and  $^{30}$SiO $J=1-0$ 
emission, established the maser nature of the emission, and confirmed  association with the deeply embedded high-mass young stellar object commonly denoted radio Source~I. 
The \te~$v=0$  maser emission shows a bipolar structure that extends over $\sim 700$~AU along a northeast-southwest  axis, and we propose that it traces a bipolar outflow driven by Source~I. 
The high-brightness isotopic SiO maser emission imaged with a $\lesssim$0\pas2 resolution has a more compact distribution, generally similar to that of the \te~$v=1,2$ emission, and it probably traces bulk gas flows in a region of diameter $\lesssim $100~AU centered on Source~I.
 On small scales of $\lesssim 10$~AU, however, compact \tn/\tr~$v=0$ and \te~$v=1,2$ emission features may be offset from one another in position and line-of-sight velocity.

From a radiative transfer  analysis based on a large velocity gradient (LVG) pumping model, we derive similar  temperatures and densities for the optimum excitation of both $^{29}$SiO/$^{30}$SiO $v=0$ and $^{28}$SiO $v=1,2$ masers, significantly higher than required for \te~$v=0$ maser excitation. In order to account for the small-scale differences among the isotopologues ($v=0$) and the main species ($v=1,2$), follow-up radiative transfer modeling  that incorporates  non-local line overlap among transitions of all SiO isotopic species may be required. 

\end{abstract}

\keywords{line: profiles - masers - ISM: individual objects: Orion BN/KL - stars: formation}

\section{Introduction}
\label{int}

The Orion Becklin-Neugebauer/Kleinman-Low (BN/KL)  region ($D \sim 414 \pm 7$~pc; \citealt{Men07}) is one of only three star forming regions (SFRs) known to exhibit maser emission from \te~\citep{Has86,Zap08} and the only one to display emission from the major SiO isotopic species comprising $^{16}$O: \te, \tn, \tr~(e.g., \citealt{God09}),
and $^{18}$O: $^{28}$Si$^{18}$O \citep{Cho05}. 

Early single-dish and interferometric observations of $J=2-1$ ($\lambda$ 3~mm) and $J=1-0$ ($\lambda$ 7~mm) rotational transitions in the $v=1$ and $v=2$ vibrational states have shown similarities among 
spectral profiles and spatial distributions, possibly indicating that they arise in the same gas volumes.
The vibrationally excited states have been known to be inverted since their first identification \citep{Tha74,Buh74}. 
 Observations with the Very Large Array (VLA)  first showed that the center of the $v=1 \ (J=1-0)$ SiO maser emission coincides with the radio continuum source denoted Source~I \citep{Chu87,Men95}. Subarcsecond imaging at near- and mid-infrared wavelengths shows no  infrared counterpart to Source~I, which suggests it is a deeply embedded young stellar object or YSO \citep{Gez92,Gre04b}. High-angular resolution imaging with Very Long Baseline Interferometry (VLBI) in both $J=1-0$ and $J=2-1$ lines has shown that the $v=1$ and $v=2$ emission traces a protostellar wind and/or outflow expanding with a characteristic velocity $\lesssim 20$~\kms at radii of 20-70\,AU from Source~I (\citealt{Gre98,Doe99,Gre04a,Doe04,Mat07}; Matthews et al. in prep.). 
 
 Thermal \te~$v=0$ emission is a well-known  tracer of shocks  in protostellar jets (e.g., \citealt{Nis07}). After the discovery of $v=0 \ J=2-1$ \citep{Buh75} and  $J=1-0$ \citep{Gen80} lines in Orion BN/KL, interferometric imaging demonstrated that the vibrational ground state emission is part maser and part thermal ($J=1-0$: \citealt{Cha95}; $J=2-1$: \citealt{Wri95}). The \te~$v=0$ traces a bipolar structure elongated northeast-southwest (NE-SW), that has been interpreted either as a rotating disk  ($J=2-1$; \citealt{Wri95}) or as a wide-angle bipolar outflow with characteristic speeds of 10-15~\kms ($J=1-0$;  \citealt{Gre04a}).
  
The \tn~and \tr~isotopologues were first discovered towards Orion BN/KL in the $v=0\ J=2-1$ line  (\citealt{Lov76}; and \citealt{Wol80}, respectively).
\citet{Olo81a} first proposed maser action for the SiO ($J=2-1$) isotopic emission, based on  observed (yearly) variability of individual Doppler components (\tn) or the asymmetric line shape (\tr).
\citet{Bau98} confirmed  maser action for the \tn~$J=2-1$ transition via high-angular resolution ($1-3''$) imaging, but there has been no conclusive evidence of maser action for the $J=1-0$ transition or for $^{30}$SiO  in either transition.
 
 The physical origin of the isotopic emission is not yet well determined. Plateau de Bure Interferometer (PdBI) imaging shows that the centroids of \tn~emission ($J=2-1$) in multiple velocity channels between $-$11 and 24~\kms arise $\lesssim $100~AU from Source I, when observed with $1-3''$ resolution \citep{Bau98}.  In contrast, the higher order $J=8-7$ transition of  \tr, observed with the Smithsonian Submillimeter Array (SMA) and comparable angular resolution, is distributed over $\sim 1000$~AU, elongated NE-SW, as is the part-thermal, part-maser \te~$v=0$ emission \citep{Beu05}.
Single-dish spectra for the ground-state \tn~and \tr~$J = 1-0$ transitions also exhibit line wings that are similar to those of \te~($\pm 40$~\kms), as recently reported by  \citet{God09}.  These are broader than the velocity extent of vibrationally excited emission and invite speculation that they might trace moderately high-velocity components   of the flows around Source\,I \citep{God09}.

Sub-arcsec resolution imaging is necessary to establish unambiguously the location and the nature (maser vs thermal)  of the isotopic emission within the BN/KL region, in particular where the high-velocity isotopic emission arises within the SiO flow.  With this purpose, we conducted {\it simultaneous} observations of the $v=0, v= 1$, and $v= 2 \ J= 1-0$ transitions of \te~and the $v=0 \ J= 1-0$ line of the \tn~and \tr~isotopologues at 7~mm wavelength with the VLA.
For the first time, we have mapped the \tn~and the \tr~ $v=0 \ J= 1-0$ lines, measured the proximity of the emission regions to Source~I, and established the presence of population inversion. Simultaneous observation of  SiO masers from three vibrationally excited transitions and  three isotopologues, provide constraints on  SiO excitation conditions in the surrounding medium. 

The  observational setup and data calibration procedures are described in \S 2.
Spectral profiles and  images of different maser transitions are presented in \S 3. In \S 4, we discuss the gas kinematics probed by  each species (\S 4.1), the presence of maser action from  \tn~and \tr~(\S 4.2), and the excitation conditions  derived from preliminary radiative transfer modeling  (\S 4.3). 
 Finally, conclusions are drawn in \S 5.

 \begin{deluxetable}{cclccc}
\tabletypesize{\footnotesize}
\tablewidth{0pc}
\tablecaption{Parameters of Observations.}
\tablehead{
\colhead{Program} & \colhead{Date} & \colhead{VLA$^{(1)}$} & \colhead{Transitions} & \colhead{$\nu_{\rm rest}$}& \colhead{$T^{(2)}$}\\ 
\colhead{} & \colhead{} & \colhead{Conf.} & \colhead{$J=1-0$} & \colhead{(MHz)} & \colhead{(hr)} 
}
\startdata
AG776 & 2008/02/02 & B (13V) & \te~$v=0$ & 43423.858 & 2.3 \\
AG776 & 2008/02/02 & B (13V) & \te~$v=1$  & 43122.080 & 4.6 \\
AG776 & 2008/02/02 & B (13E) & \te~$v=2$  & 42820.587 & 4.6 \\
AG776 & 2008/02/02 & B (13V) & \tn~$v=0$ & 42879.846 & 2.3  \\ 
AG776 & 2008/02/02 & B (13E) & \tr~$v=0$ &42373.340 & 4.6 \\
AG575 & 2000/01/01& B (19)& \te~$v=1$  & 43122.080 & 0.5 \\
AG575 & 2000/01/01& B (19)&\tn~$v=0$ & 42879.846&0.5   \\ 
AG575 & 1999/08/28& A (10) & \te~$v=1$  & 43122.080 & 0.8   \\
AG575 & 1999/08/28& A (10)&\tn~$v=0$ & 42879.846 & 0.4 \\ 
\enddata
\tablecomments{\\
\scriptsize 
$^1$ Spacing of antennas: B-configuration has baselines of 0.1 to 11 km.  A-configuration has baselines of 0.5 to 35 km.  The number of antennas participating is in parentheses, trailed by ``V" to indicate VLA and ``E" to indicate EVLA.  Prior to 2008, the number of participating antennas corresponds to the number equipped with high-frequency receivers.  All setups placed either $v=1$ or $v=2$ $^{28}$SiO in one IF to enable high-accuracy  calibration.\\
$^2$ Approximate on-source integration time. 
}
\label{obs}
\end{deluxetable}

\section{Observations and data reduction}
\subsection{Observations}
Observations of the SiO emission from Source~I were conducted at 7~mm 
using the Very Large Array (VLA) of the National Radio Astronomy
Observatory (NRAO)\footnote{NRAO is a facility of the National Science Foundation
operated under cooperative agreement by Associated Universities, Inc.}.
We present data obtained on 2008 February 2, along
with archival data from 1999 August 28 and 2000 January 1.  
The data from 2000 and 2008 were obtained while the VLA was in the
B-configuration, yielding a resolution of $\sim$0\pas2.
The data from 1999 are from the A-configuration, giving a
resolution of $\sim$0\pas5.  Table~\ref{obs} summarizes the observational
parameters of the different programs.  The following description refers
to the program observed in 2008.

We observed a total of five $J=1-0$ rotational transitions from three
different SiO isotopologues: 
$^{28}$SiO $v=0,1,2$; $^{29}$SiO $v=0$;
and $^{30}$SiO $v=0$ (Table~\ref{obs}).
To achieve high-accuracy astrometry, we cycled our observations 
among the various transitions, observing two at a time.  
We included emission from the $v=1$ or $v=2$ states of \te~in each 
of the setups because these were strong enough to be used to calibrate 
the troposphere and system gain, thus facilitating detection of the 
weaker emission from the other isotopologues.
At the time of the 2008 observations the VLA included 13 
EVLA\footnote{The Expanded Very Large Array is a joint USA, Mexico, 
and Canada project to greatly improve the observing capabilities of 
the VLA array by installing new electronics (fiber optic cables, 
large bandwidth receivers, new correlator, etc.) while keeping the 
existing infrastructure (antennas, railroad tracks, etc.).} 
antennas that were able to simultaneously observe two separate 
bandpasses (IFs) up to 2 GHz apart. In contrast, the maximum
separation between IFs allowable with the remaining (VLA) antennas
was $\sim400$~MHz, depending on details of the available 
local oscillator settings.
In order to optimize the use of the VLA and EVLA antennas,
respectively, we therefore divided 
the array into two
sub-arrays of 13 antennas each. The VLA antennas observed the
\te~$v=1$ maser at 43122 MHz in one IF (single polarization) and the
\te~and \tn~$v=0$ lines at 43424/42880 MHz alternately in the other IF
(single polarization). The EVLA antennas observed the \te~$v=2$ maser
at 42821 MHz simultaneously with the \tr~$v=0$ line at 42373 MHz (all single
polarization). 
For \tr~$v=0$, a tuning offset of 2.83~MHz was
introduced alternately with the main tuning to enable a search for possible
high-velocity emission (up to 70~km~s$^{-1}$), whose presence was 
suggested in earlier single-dish spectra \citep{God09}. No such
emission was detected by the present observations. 
The total on-source integration times for the various SiO 
transitions are given in Table~\ref{obs}, and 
the frequency pairings used in the observing programs 
are summarized in Table~\ref{pairs}.
The \te, \tn, and \tr~$v=0$ transitions
were each observed with a 12.5\,MHz total bandwidth (88\,km\,s$^{-1}$) with a
spectral resolution of 390.6~kHz (2.7\,km\,s$^{-1}$).  The \te~$v=1,2$
emission was observed with a 6.25\,MHz bandwidth and 97.656\,kHz
(0.68\,km\,s$^{-1}$) channels (Table~\ref{pairs}).

We added several continuum-mode setups to the 2008 observations to
estimate any frequency dependent (i.e., delay) terms to the astrometry
error budget that 
might affect the relative astrometry between the 
pairings of transitions observed in spectral-line mode.  
In particular, we performed 
observations of the respective \te\ maser ($v=1$ for the VLA subarray, 
$v=2$ for the EVLA subarray) in one IF 
(single channel, 6.25 MHz bandwidth, dual polarization) together 
with observations in a 50 MHz continuum channel in the other IF 
(also dual polarization), offset in frequency by $\pm$350 MHz from the
maser. The goal was to detect and image the BN object and use its 
position to align the various SiO emissions.  
 Unfortunately these observations on the EVLA subarray failed, 
so we have aligned the $v=1$ and $v=2$ emissions with respect to 
each other using other techniques (Section~\ref{alig_abs}).

We derived  absolute flux calibration from observations of 3C~286
($F_{\nu}= 1.5$~Jy), and bandpass calibration  from
observations of 3C~84.  The phase offsets between the paired IFs were tracked
through observations of QSO J0541$-$0541 every 25 minutes.  The
strong \te~$v=1$ and $v=2$ masers ($\gg 100$~Jy) were used to track 
tropospheric phase
fluctuations on 10-s timescales, as described in Section~\ref{cal_pro}.
All spectral-line observations were observed in a fixed-frequency mode,
and Doppler corrections were applied during the data reduction.

Observations of \tn~$v=0$ and \te~$v=1$ in 1999 and 2000
(Table~\ref{obs}) were made with setups similar to those used in 2008
for the VLA antennas.  
We adopted a 2-IF, single polarization mode, with bandwidths of 
6.25~MHz.  Monitoring of phase offsets between the IFs was accomplished 
with periodic observations of QSO J0541$-$0541, and bandpass
calibration relied upon observations of QSO J0530+135.
 %
 \begin{deluxetable*}{clcccccccccc}
\tabletypesize{\footnotesize}
\tablewidth{0pc}
\tablecaption{Frequency pairs in the observing setup}
\tablehead{
\colhead{Prog.} & \colhead{Mode} &&& \colhead{IFA(-C)}&\colhead{BW} & \colhead{CW} &&& \colhead{IFB(-D)}&\colhead{BW} & \colhead{CW}  \\ 
\colhead{} & \colhead{} & \colhead{} && &\colhead{(MHz)} & \colhead{(kHz)} & \colhead{ } && &\colhead{(MHz)} & \colhead{(kHz)} 
}
\startdata
AG776 & Line&&&\te~$v=0$ & 12.5& 390.6 & &&\te~$v=1$ & 6.25& 97.7 \\
AG776 & Line&&&\tn~$v=0$ & 12.5 & 390.6&& &\te~$v=1$ & 6.25& 97.7\\
AG776 &Line&&& \tr~$v=0$& 12.5 & 390.6& &&\te~$v=2$ &6.25& 97.7 \\
AG776 & Cont. &&& $v=1 +350$& 50.0 &-&&& \te~$v=1$& 6.25&-\\
AG776 & Cont. &&& $v=1 -350$& 50.0 & -&&&\te~$v=1$& 6.25 & -\\
AG776 & Cont. &&& $v=2 +350$ & 50.0 &-& &&\te~$v=2$ & 6.25& -\\
AG776 & Cont. && &$v=2 -350$ & 50.0 &-&& &\te~$v=2$ & 6.25&- \\
AG575 &Line &&& \tn~$v=0$ & 6.25 & 97.7 &&& \te~$v=1$ & 6.25& 97.7 \\
\enddata
\tablecomments{
\scriptsize 
Cont. = Continuum and Line = Spectral-line mode: 64 (32) channels of 97.7 (390.6)~KHz width (CW)  within a 6.25 (12.5)~MHz bandwidth (BW).
}
\label{pairs}
\end{deluxetable*}
%
\subsection{Calibration procedures}
\label{cal_pro}

Calibration and data reduction were carried out using the Astronomical
Image Processing System (AIPS).  
First,  the phase offset between the two simultaneously-observed  
IFs due to the electronics was removed, so that subsequent
tropospheric 
calibration from the \te\ $v=1$ and $v=2$ transitions could be applied 
directly to the IF containing the weaker line.  The phase offset was determined
from observations of J0541$-$0541 in the following manner.  Using the
12.5 MHz bandwidth IF, calibration solutions were determined for the
short-term tropospheric phase fluctuations on J0541$-$0541 and applied
to J0541$-$0541 in the 6.25 MHz IF.  A subsequent phase calibration on
the 6.25 MHz IF using J0541$-$0541 and averaging over scan lengths 
($\sim 2$~min)
then gave the phase offset of this IF relative to the 12.5 MHz, and was
applied to all sources in the dataset.  Amplitude calibration was subsequently derived
from J0541$-$0541, after correcting for the short-term tropospheric phase fluctuations. We also derived separately  phase solutions from 2-minute scan averages of J0541$-$0541 and applied them to Source I. 
Corrections for the shape of the antenna-based bandpasses were also applied, along with Doppler corrections.

Finally, corrections for fluctuations in the tropospheric and instrumental
phase and amplitude were derived every 10 seconds from channel averages
of the strong \te\ masers ($V_{\rm LSR} = -7.2$ to $-$3.1 km~s$^{-1}$
for the $v=1$ line, and $-$9.4 to $-$5.3 km~s$^{-1}$ for the $v=2$ line)
using self-calibration and applied to the IF containing the weaker line.
As a consequence, maps of the ground-state transitions are (independently)
referenced to the positions of the  \te~$v=1$ or $v=2$ emission.  

Images were made of Source~I using the AIPS task IMAGR with a 
``robust'' parameter of 0, resulting in a synthesized beam size of
$\sim$0\pas2 for the B configuration data and $\sim$0\pas05 for the 
A configuration data. The images were 1024 pixels on a side with 
cell sizes 0\pas02 (2008) or 0\pas005 (1999) and covered
fields-of-view of 
20$''$ and 5$''$, respectively.
For the \te~images the sensitivity is dynamic range
limited by the sidelobes of the strongest emission peaks and thus
varies among different frequency channels and different transitions.
Table~\ref{results} summarizes the imaging results.

\begin{deluxetable*}{cccccccc}
\tabletypesize{\scriptsize}
\tablewidth{0pc}
\tablecaption{Summary of Observational Results}
\tablehead{
\colhead{Date} &  \colhead{Transitions} & \colhead{Synthesized Beam$^{(1)}$} & \colhead{RMS$^{(2)}$} & \multicolumn{2}{c}{Blue Peak$^{(3)}$}   & \multicolumn{2}{c}{Red Peak$^{(3)}$}  \\
\colhead{} & \colhead{$J=1-0$} & \colhead{$\theta_M('') \times \theta_m(''); \ P.A.(^{\circ})$} & \colhead{(mJy~beam$^{-1}$)} & \colhead{V(km~s$^{-1}$)} & \colhead{I(Jy~beam$^{-1}$)} 
& \colhead{V(km~s$^{-1}$)} & \colhead{I(Jy~beam$^{-1}$)}  
}
\startdata
AG776 &  \te~$v=0$  & $0.23 \times 0.19; 34$ &7--113& $-5.8$ & 10 & 13.1& 4.9   \\
&\te~$v=1$  &  $0.23 \times 0.19; 33$ & 15--170 & $-4.5$ & 1590& 14.5 & 930\\
& \te~$v=2$  &  $0.24 \times 0.15; 20$ &6--40& $-6.6$ &470 &20.0  & 380 \\
& \tn~$v=0$ &  $0.22 \times 0.19; 25$ & 5 & $-3.7$ & 0.044 & 15.5 & 0.12\\ 
& \tr~$v=0$ & $0.25 \times 0.15; 21$ & 5 & $-4.6$ & 0.068 & 14.8 & 0.072 \\
AG575& \te~$v=1$  & $0.20 \times 0.17; 35$ & 11--2003 & $-3.2$ & 340& 14.5 & 210\\
& \tn~$v=0$ &  $0.20 \times 0.17; 7$ & 7--27 & ... & $<0.05$ & 16.2 & 0.34\\ 
AG575 &  \te~$v=1$   & $0.061 \times 0.045; 53$ & 50-200 & $-3.9$ & 600& 13.8 & 400\\
& \tn~$v=0$ &  $0.059 \times 0.042; 45$ & 16 & ... & $<0.07$ & 15.9 & 0.840\\ 
\enddata
\tablecomments{\\
\scriptsize 
$^{(1)}$ The synthesized beam results from a robust 0 weighting. \\
$^{(2)}$ $1\sigma$ RMS noise in a 0.7 or 2.7~\kms channel,  based on histogram fits to the pixels over the channel image. \\
$^{(3)}$ Velocity and intensity of the emission peak on the blue and red side of the spectra.
}
\label{results}
\end{deluxetable*}
%
%
%
\subsection{Positional registration among different transitions}
\label{alig_abs}
Phase and amplitude self-calibration enable coherent integration and imaging but residual astrometric errors remain, due to the $\sim 1^{\circ}$ separation between calibrator (J0541-0541) and target.
To remove frequency and sky position dependent errors (i.e., the residual delay error due to the  atmosphere), we used as a position reference for registration the peak of the continuum emission from BN, which is the strongest in Orion BN/KL ($F_{\nu} \sim 25$~mJy at 7~mm) and only $\sim 10$\arcsec~from  Source I. 
For that purpose, we calibrated the continuum-mode dataset from 2008 following a procedure similar to that described in Sect.~\ref{cal_pro} for the line-mode dataset.
The band-averaged signal-to-noise ratio (SNR) for the $v=1$ and $v=2$ maser emission over 6.25 MHz was high enough that the narrow band data enabled coherent integration and detection of BN in the broad band. 

By directly comparing the positions of BN in bands tuned to the $v=1$ emission and offset by $\pm 350$~MHz, we established  the angular offset per MHz introduced by calibrating one band by another: $\Delta \alpha = 0.007 \pm 0.003$~mas/MHz and  $\Delta \delta =  -0.016\pm 0.004$~mas/MHz.
Unfortunately, the EVLA subarray tuned to the $v=2$ emission did not provide usable data in the continuum mode setups. We used the   offset derived with the VLA subarray to  correct the positions of all pairs of transitions with 1 mas accuracy.

Since the $v=1$ and $v=2$ transitions served as calibration references for different subarrays and were self-calibrated independently, without an EVLA detection of BN it is necessary to register them separately to each other. For that purpose we measured the intrinsic structure of the $v=1,2$ emission using VLBA images from a  monthly monitoring of Source~I over several years (\citealt{Mat07}; Matthews et al. in prep.).
After averaging in velocity to account for  different spectral resolution, the VLBA images were convolved with the VLA beam, and the offset between the $v=1$ and $v=2$ emission was measured. 
We did tests using 4 to 5 different velocity ranges in the $v=1,2$ spectra at 4 to 5 different epochs from the VLBA monitoring and obtained consistent results with an offset of $\Delta \alpha = 14 \pm 3$ and  $\Delta \delta =  0.5 \pm 3$~mas.
The accuracy in the computed offsets is dominated by the variation of the structure and of the relative locations of the $v=1$ and $v=2$ maser centroids from channel to channel and from epoch to epoch, rather than by the uncertainties in measuring the locations of the $v=1,2$ emission peaks.  The quoted uncertainty is then the variance of the offset values measured in several channels and at different epochs. 

After applying  the angular offset per MHz relation derived above and the offset between the $v=1$ and $v=2$ emission,  we estimate that the maps of the various transitions are  aligned on the sky with accuracy of 3 mas in each direction.  

\subsection{Absolute astrometry}
\label{astro}
The application of the phase solutions derived from 2-minute scan averages of J0541$-$0541  to the maser data imposes an absolute astrometry  (Section~\ref{cal_pro}).  However, because J0541$-$0541 was observed relatively
infrequently (every 25 minutes) compared with the timescale of
tropospheric and instrumental phase variations, the absolute astrometry
is poor, and further alignment is necessary. In this case, 
absolute positions were established by comparing positions of  BN in our maps with the position derived  by \citet{Gom08} from VLA observations conducted in May 2006 at 8.4~GHz, $\alpha(J2000) = 05^h 35^m 14\rlap{.}^s1099$,   $\delta(J2000) = -05^{\circ} 22' 22$\pas741 (positional accuracy $\sim $0\pas01) 
and their estimated proper motion ($V_{\rm x}=-5.3$~mas yr$^{-1}$, $V_{\rm y}=9.4$~mas yr$^{-1}$). For 2008 February, we obtain 
$\alpha(J2000) = 05^h 35^m 14\rlap{.}^s1093$,   $\delta(J2000) = -05^{\circ} 22' 22$\pas7253. 
 The offset between the position of BN measured in our maps and  and the   proper-motion-corrected position from  \citet{Gom08}
  was applied to the SiO images, whose absolute position is then measured to an accuracy of 0\pas01. 
  

\subsection{Fitting positional centroids of maser emission} 
\label{ide}

The  emission in each velocity channel is unresolved in all transitions except the \te~$v=0$. For the latter, in order to establish the morphology and kinematics of the extended emission, we produced an intensity map integrated over the emission velocity range. 
For each unresolved emission component, a two-dimensional ellipsoidal Gaussian model is fitted in a $100 \times 100$ pixels   area of each channel map.  Since the angular extent of the maser emission in each velocity channel is smaller than the synthesized beams of the VLA in the observing configurations, spatially separated maser components at similar velocities are blended together.  Hence, the Gaussian-fitted positions  actually locate the intensity-weighted centroids of SiO emission in each frequency channel. In the following, we will refer to these emission centroids in individual velocity channels as maser {\it features}. 

 The relative positional errors of maser features are limited  by the SNR of the channel maps and by systematic errors in the calibration of the bandpass. In the case of weak signals (e.g., \tn~and \tr), the positional uncertainties are noise limited, and are given by
$\delta\theta = 0.5 \theta_B/{\rm SNR}$, where $\theta_B$ is the
FWHM of the synthesized beam, and SNR is the peak intensity divided
by the RMS noise in a particular velocity channel.
 For the 2008 maps ($\theta_B=$0\pas2),  the relative positional uncertainties are typically $\sim$~10~mas for \tn~and \tr~ ($\sim $50~mJy~beam$^{-1}$ peak and $\sim$5~mJy~beam$^{-1}$ RMS), whereas for the 1999 maps ($\theta_B=\sim 50$~mas), the typical uncertainties are $\sim 1$~mas for \tn~($\sim $500~mJy~beam$^{-1}$ peak and $\sim$17~mJy~beam$^{-1}$ RMS).
  For the $v=1,2$ emission, the positional uncertainties are dominated by frequency-dependent errors in calibrating the bandpass. From the continuum spectrum of 3C~84, we measure a channel to channel phase noise $<1^{\circ}$, which corresponds to  an uncertainty of $<0.6$~mas for a 0\pas2 synthesized beam.

\begin{figure}
\centering
\includegraphics[width=0.5\textwidth]{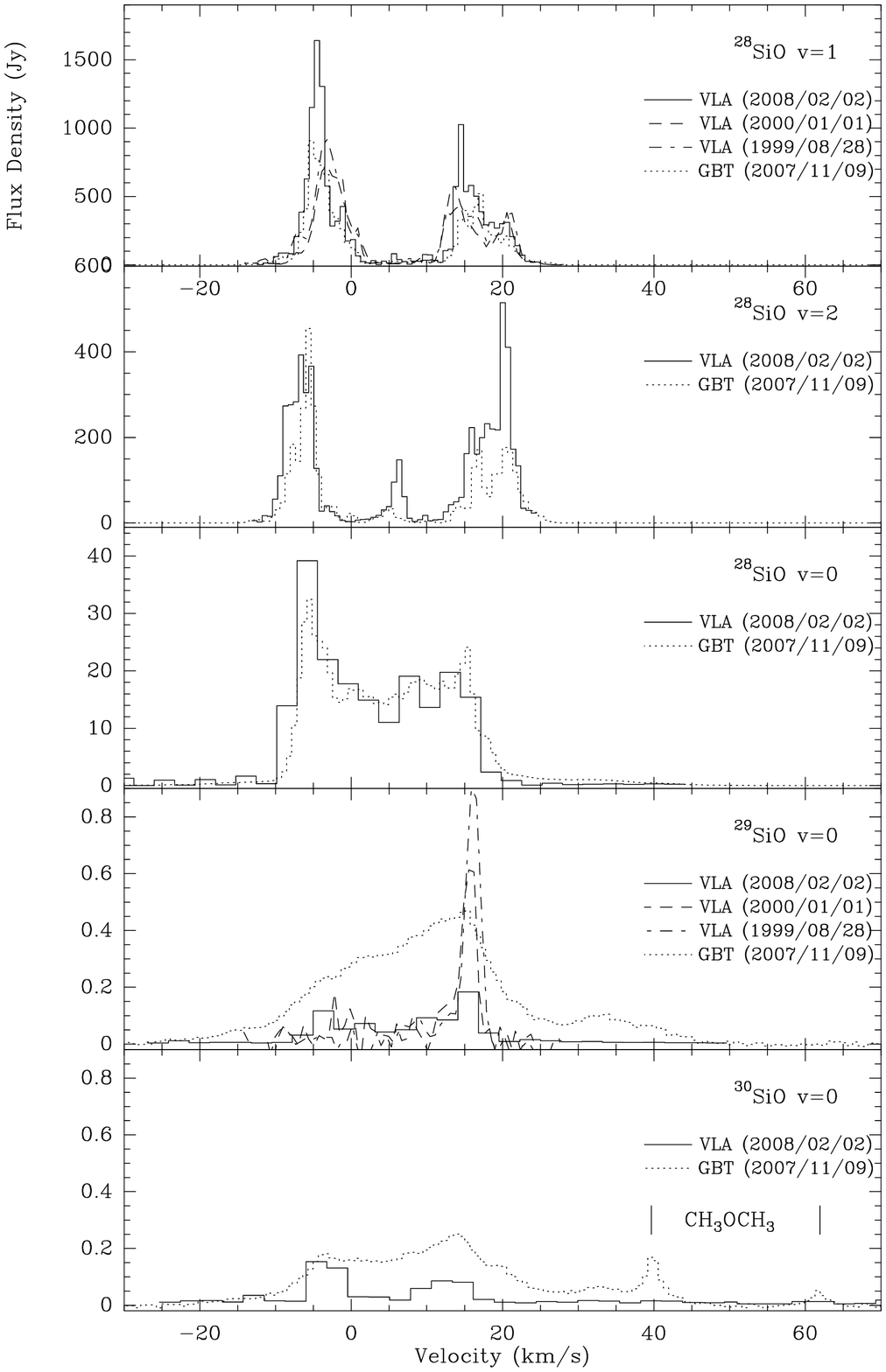}
\caption{Spectra of various SiO transitions observed toward Orion~BN/KL   with the VLA  during several epochs. GBT spectra from \citet{God09} are also shown for comparison.
In 2008, the VLA spectral resolution was 97.7~kHz (0.68~\kms) for the \te~$ v=1,2 \ J=1-0$ lines and 390.6~KHz (2.7~\kms) for the \te, \tn, and \tr~$v = 0 \ J = 1-0$ lines. In 1999 and 2000, \te~$ v=1$ and \tn~$v=0$ were observed with a 0.68~\kms resolution. The large differences between GBT and VLA $v=1,2$ spectra is likely to be due to the time separation of observations and the strongly maser-nature and compactness of the emission. In the case of the isotopologues, most probably the observed difference is due to an extended, thermal component detected by the single-dish but resolved out by the interferometer.
Note that the same flux density scale is adopted for \tn~and \tr.
}
\label{spec}
\end{figure}
\section{Results}
\label{res}

We imaged the \te~$v=0,1,2$ and the \tn~and \tr~$v=0 \ J=1-0$ transitions observed in 2008 (0\pas2 beam), as well as the \te~$v=1$ and   \tn~$v=0$ transitions observed in 2000 (0\pas2 beam) and 1999 (0\pas05 beam).

In each epoch and for each transition, we produced spectra by mapping each spectral channel (0.68 or 2.7~\kms wide) and then summing up the flux density in each channel map.
Figure~\ref{spec} compares for each transition the integrated flux densities of the VLA channel maps at different epochs with the GBT single-dish spectra (observed in 2007  November 10: \citealt{God09}). 
Multiple transitions show similar double-peaked profiles and velocity extents: blueshifted emission in the velocity range from $-10$ to 0~\kms, redshifted emission in the interval from 10 to 25~\kms , and systemic emission in the range from 0 to 10~\kms.
Comparison of  the GBT (2007 November) and the VLA (2008 February) spectra, shows an increase in  the  flux density of about 100\%   in 
the emission of $v=1$ (both blue and red components) and $v=2$ (only red component), whereas the \te~$v=0$ emission  does not show any significant variation. 
Conversely,  less than $\sim 30$\% 
of the single-dish emission has been detected in the VLA channel maps for \tn~and \tr. 
The single-dish spectral profiles of the latter two transitions appear much broader than the VLA profiles and show high-velocity wings  undetected in the VLA channel maps (Fig.~\ref{spec}).
 Interestingly, the \tn~spectrum has undergone significant changes over 8-9 years (the separation between VLA epochs). In the spectra taken in 1999 and 2000, the redshifted component  is 10 and 5 times stronger, respectively, than in the 2008 spectrum while no emission from the blue component is detected above a $4\sigma \sim 50-70$~mJy noise level (Fig.~\ref{spec}). 

\begin{figure*}
\centering
\includegraphics[angle= -90,width=\textwidth]{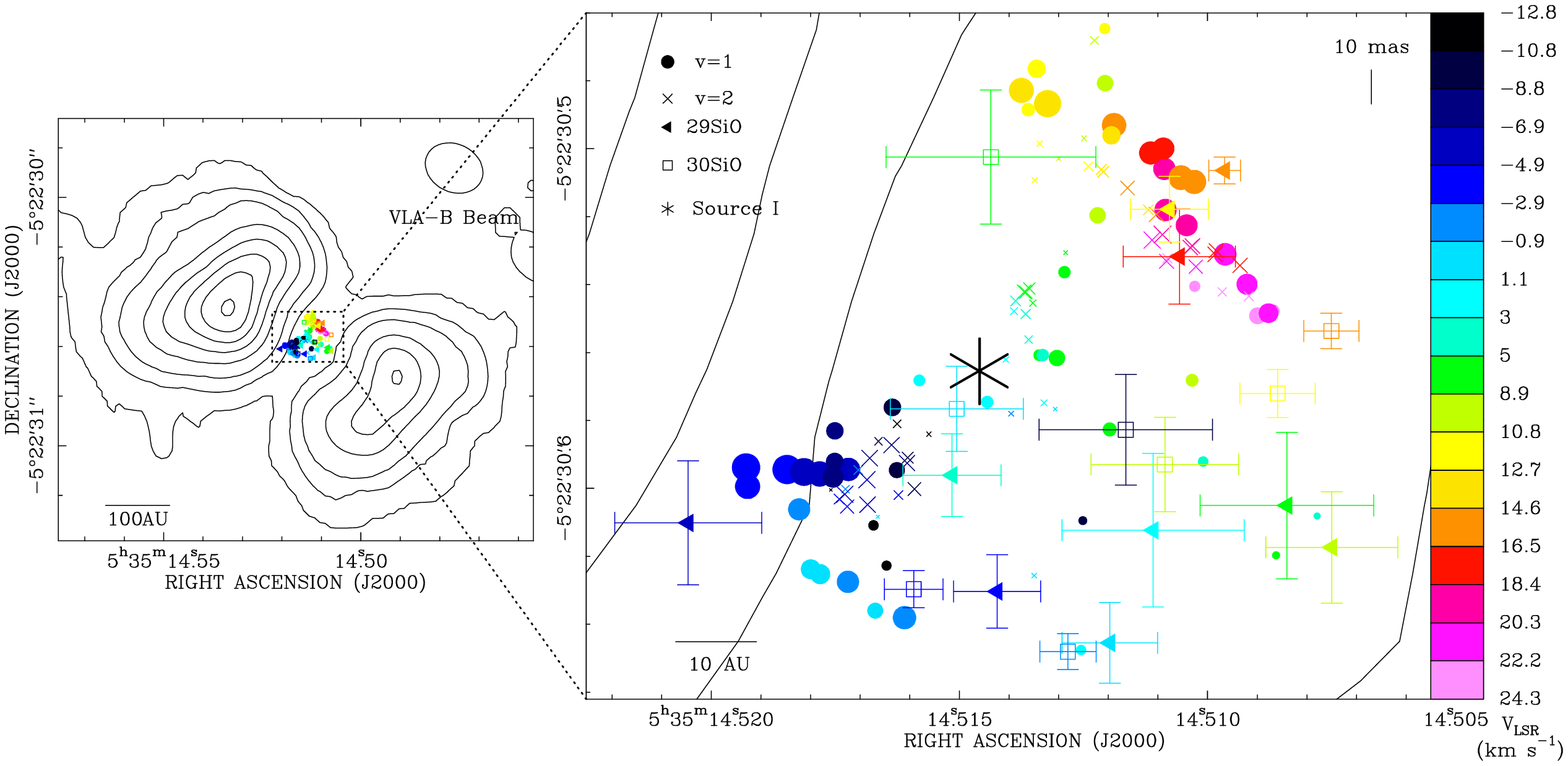}
\caption{Distribution of isotopic SiO maser emission in Orion BN/KL observed on 2008 February 2.
({\it Left}) Emission map of  \te~$v=0 \ J=1-0$ is integrated    from -11 to 21~\kms. The contour levels are 1, 5, 10,  20,  30,  40,  50, 60 Jy~beam$^{-1}$~\kms. The map shows a bow tie shape similar to that observed previously at 3~mm wavelength with BIMA \citep{Wri95}. Source~I is located at the center of the \te~$v=0$ flow, where the emission from other isotopologues and vibrational states also lie. The beam (0\pas22 $\times$ 0\pas19, P.A. 25$^{\circ}$) is shown on the right upper corner. ({\it Right}) Emission centroids of \te~$v=1$ ({\it filled circles}) and  $v=2$ ({\it crosses}), \tn~({\it filled triangles}) and \tr~$v=0$ ({\it open squares})  as a function of velocity superimposed on the velocity integrated intensity map of \te~$v=0$ ({\it contours}). 
 {\em Color} denotes  L.O.S.  velocity (color scale on the right-hand side). The sizes of circles and crosses  scale logarithmically with the peak intensity of the $v=1,2$ emission. For \tn~and \tr~the position uncertainties reflect formal errors in model fits. The adopted absolute  position of Source I is 
$\alpha(J2000) = 05^h 35^m 14\rlap{.}^s5146\pm0\rlap{.}^s0007$,   $\delta(J2000) = -05^{\circ} 22' 30$\pas5655$\pm0$\pas01, after correction for proper motions  \citep{Gom08}; the size of the  {\it star} indicates uncertainty. The linear and angular spatial scales are given in the lower left and upper right corners of the panel, respectively. The relative alignment between pairs \te~$v=1$ and \te/\tn~$v=0$ or \te~$v=2$ and \tr~$v=0$ is accurate to $\sim$3~mas. The uncertainty in the absolute position is $\sim$10~mas.
}
\label{ag776}
\end{figure*}

The velocity integrated intensity map of \te~$v=0$ emission consists  of  two bright peaks offset  $\sim \pm$0\pas5 from Source~I  (Fig.~\ref{ag776}, left panel). The bipolar structure  extends over $\sim 700$~AU along a NE-SW direction and is similar to previous BIMA images of the $J=2-1$ line \citep{Wri95}.
Both lobes emit in a similar velocity interval, from --9 to 16~\kms.

The emission centroids from  \te~$v=1,2$ and \tn~and \tr~$v=0$ are clearly located at the center of the \te~$v=0$ bipolar structure (Fig.~\ref{ag776}, left panel) and are distributed in a region of size 0\pas2 around Source~I (Fig.~\ref{ag776}, right panel).
They show  generally similar spatial and velocity distributions and can be divided into three groups:  {\it blue features} detected southeast (SE) of Source~I include emission with line-of-sight (L.O.S.) velocities blueshifted with respect to the systemic LSR velocity of the region (5~km~s$^{-1}$), {\it red features} located to the north-west (NW) have emission with redshifted L.O.S. velocities, and {\it systemic features}, which are weaker and located in between the two other groups, are comparatively close to  the systemic velocity. 
However, some differences can be noted among various transitions.
Compared to $v=1$, the \te~$v=2$ emission appears located in a slightly inner layer, closer to Source~I, even if there is a significant overlap. In addition, the systemic features appear stronger (relative to blue and red peaks) for $v=2$ than for $v=1$. This is consistent with the findings of higher-resolution imaging of the \te~$v=1,2$ emission with the VLBA  (\citealt{Mat07}; Matthews et al. in prep.). 
 
As far as the isotopologues are concerned, although correspondence among ground-state isotopologue and vibrationally excited emission distributions is generally good, a one-to-one correspondence in positions and L.O.S. velocities in  features among different transitions  is clearly absent. 
 In addition,  the brightest features from the main species and isotopologues are not co-spatial (with the exception of the \tn~red features in the NW area). 
Finally, while most of the blue and systemic \tn~features also have corresponding \tr\ features, this is not necessarily true for the red features.

\begin{figure}
\centering
\includegraphics[width=0.5\textwidth, trim = 0cm 2.3cm 0cm 3.2cm,clip]{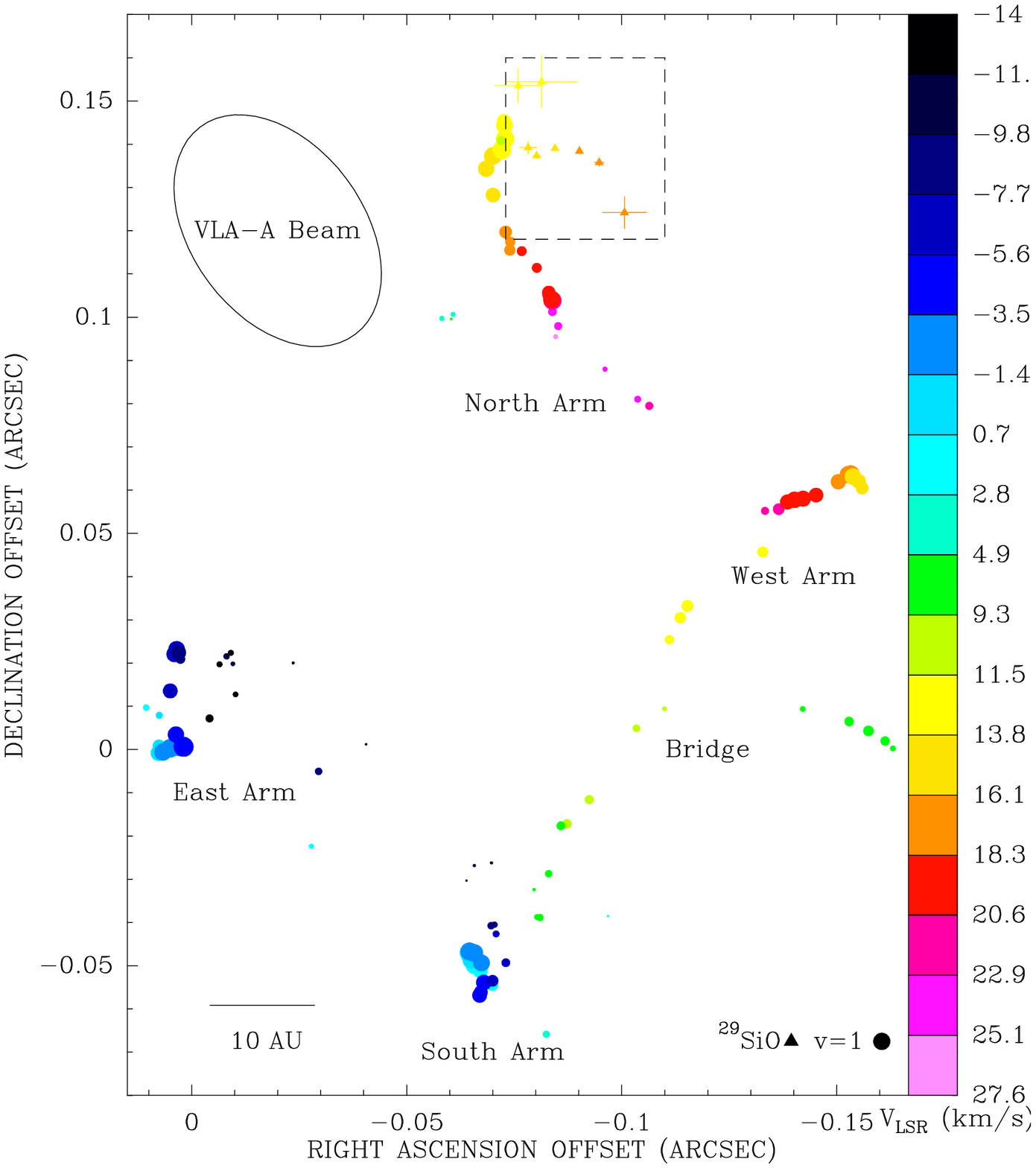}
\caption{ Centroids of  \te~$v=1$ ({\it filled circles}) and \tn~$v=0$  ({\it filled triangles}  with error bars)  compact emission observed on 1999 August 28. The \te~$v=1$ emission distribution traces a 4-arm structure, redshifted in the north and west arms and blueshifted in the south and east arms. Note also a weaker bridge of emission connecting west and south arms. The \tn~emission is concentrated towards the north arm ({\it dashed rectangular area}). 
The ellipse on the top left indicates the synthesized beam (0\pas06 $\times$ 0\pas04; P.A. 53$^{\circ}$). Since the beam size is comparable to the arm extensions, the $v=1$ emission from each arm is not resolved in the map, and the specific structure within each arm may be artifactual owing to spatial blending (compare with the VLBA images from \citealt{Gre98}). 
Note that the same physical scale is used in both Fig.~\ref{ag776} and Fig.~\ref{ag575}.
}
\label{ag575}
\end{figure}

We caution that, owing to low angular and spectral resolution,  blending may hamper detailed comparisons of emission centroid positions.   
For instance,  higher angular (0\pas05 beam) and spectral (0.68~\kms) resolution imaging with the VLA reveals that the \te~$v=1$ maser emission does not arise from two (blue and red) arcs (as shown in Fig.~\ref{ag776}), but resolves into an ``X"-shaped pattern, consisting of four ``arms", extending towards north (N), west (W), south (S), and east (E), plus a bridge of emission joining the S and W arms of the ``X"  (Figure~\ref{ag575}). 
     The  X-shaped structure  is consistent with sub-milliarcsecond resolution VLBA imaging  (\citealt{Gre98,Doe04}; Matthews et al. in prep.).   
In the high-resolution VLA map,  most of the red features arise  from equally bright components  in the N and  W arms, the blue features from components in the S and E arms, and the systemic features from the bridge (Figure~\ref{ag575}).  
 As a consequence, with a 0\pas2 angular resolution and 2.7~\kms spectral resolution, centroids of both red and blue features fall between N and W arms and S and E arms, respectively, resulting in the two (red and blue) arcs  seen in Fig.~\ref{ag776}.
 Similar arc morphologies have been previously seen in low-resolution imaging at 7~mm with the VLA (0\pas2~beam - \citealt{Men95}) and at 3~mm with BIMA (0\pas5 beam - \citealt{Wri95}) and PdBI (2$''$ beam - \citealt{Bau98}).  
  
 The high resolution VLA image allows  accurate localization of the \tn~emission towards the N arm of the X traced by the $v=1$ emission (Fig.~\ref{ag575}), which invites speculation that the presence of isotopologue emission within the arcs on either side of Source~I (Fig.~\ref{ag776}), when observed at lower resolution, is indicative of an origin in the four arms.  Nonetheless, detailed comparison requires caution in light of the apparent $2-10$~AU offset between \tn\ and \te\ features of similar velocity (Fig.~\ref{ag575}).  

%

\section{Discussion}

 \subsection{Gas kinematics around Source~I}
\label{maps}

The NE/SW-oriented bipolar structure traced by the \te~vibrational ground state has been previously  interpreted either as a rotating disk, based on maps of the $J=2-1$ emission at 3~mm wavelength \citep{Wri95}, or as a bipolar outflow, based on the distributions of hotspots of $J=1-0$ emission at 7~mm  \citep{Gre04a} and the distribution of the $J=8-7$ emission at 0.9~mm \citep{Beu05}.
The data presented here are consistent with the outflow hypothesis, where we cite: (i) a fully developed bipolar geometry when imaging detects both hotspots \citep{Gre04a} and extended structure (this work); (ii) lobes that exhibit similar mixes of blue and redshifted components, indicative of outflow close to the plane of the  sky rather than rotation; and (iii) the putative outflow in the vicinity of Source I is aligned with the principle axis of the ``18~\kms" outflow in the BN/KL region \citep{Gen81}, which is suggestive of Source I as a possible contributor.

The  X-shaped distribution of vibrationally excited \te~emission has been interpreted as tracing either a biconical outflow with a SE-NW axis \citep{Gre98,Doe99}, funnel-like flows arising from the  surface of a disk with a NE-SW rotation axis \citep{Gre04a}, or the interface between infalling and outflowing material \citep{Cun05}. 

Emission from both vibrationally excited states of the  main species and the ground state of the isotopologues  traces the innermost ($\lesssim $100~AU) circumstellar gas around Source~I  (Fig.~\ref{ag776} and~\ref{ag575}).
Notwithstanding similarities, it is not however possible to make  an exact position-velocity-intensity pairing of the \te~$v=1,2$ and \tn/\tr~$v=0$ features (Sect.~\ref{res}). 
Hence, it remains possible that the main and isotopic species might be excited in somewhat different volumes and in principle might require slightly different physical conditions (see  Sections~\ref{lvg}). A corresponding uncertainty in interpretation has been raised  by  \citet{Bau98} to explain differences in position between  the \te~$v=1$ and \tn~$v=0 \ J=2-1$ emission in 3~mm wavelength PdBI images. 

Previous imaging of {\it thermal} emission from \te~and \tr~($v=0 \ J=8-7$) in the submm-wave band  with the SMA  has shown that both \te~and \tr~emissions subtend similar angles on the sky (size$\sim 1000$~AU), are elongated NE-SW, and cover about the same velocity range \citep{Beu05}.  The natural  inference is that both species may trace inner reaches of the NE-SW outflow. Our VLA maps, however, do not show any indication of  isotopic emission within the NE-SW outflow traced by \te~$v=0$ emission. 
The actual spatial structure of the \tn~and \tr~$v=0 \ J=1-0$ emission  might be  more complex and extended than illustrated in Figures~\ref{ag776} and~\ref{ag575}. 
The large difference in flux between the single-dish and the interferometer spectra (Fig.~\ref{spec}), may be in fact ascribed to an extended, low-surface brightness component, detected by the GBT but filtered out by the VLA.
 Indeed, in  our VLA maps  we are insensitive to emission with an integrated flux density $\lesssim 100$~mJy and that extends over $\gtrsim 1''$ (note that the line wings observed in the GBT spectra for \tn~and \tr~are $\sim$100~mJy).
 Hence, the \tn~and \tr~emission imaged with the VLA traces only the highest brightness regions in the proximity of Source~I. 
With the present VLA data, we cannot establish the location of the moderately high-velocity component of the protostellar outflow from Source~I, as traced by the $v=0$ wing emission observed with the GBT. 


\subsection{Nature of the isotopologue emission}
\label{nat}
The \tn~and \tr~$v= 0 \ J = 1-0$ emissions have been imaged at sub-arcsecond resolution for the first time. \citet{Cha95} showed that the \te~$v=0 \ J=1-0$ emission in Orion BN/KL is part thermal and part maser ($T_b \sim 10^5$~K). \citet{Bau98} reached the same conclusion for the \tn~$v=0 \ J=2-1$ emission ($T_b \sim 10^3$~K).

On one hand, the broad profiles and the large flux densities at systemic velocities in the ground-state emission as compared to the vibrationally excited states (Fig.~\ref{spec}) clearly indicates thermal emission from all species. 
On the other hand, the  high flux density measured for the strongest features in the interferometric maps suggests, similarly to \te, non-thermal emission for \tn~and \tr~$v=0 \ J=1-0$ as well. Although our observations do not resolve the  emission in individual velocity channels (and hence we do not know the actual size of individual features), we may use the synthesized beamwidth and the observed peak flux density at different angular resolutions to estimate a minimum brightness temperature. 
For \tn, the peak intensity measured in the 16~\kms channel map with a 0\pas2 resolution  is 0.12~Jy~beam$^{-1}$, which corresponds to a line brightness temperature of $>1900$~K. 
In the 1999 observations, with an angular resolution of 0\pas05, the same velocity component has a peak intensity of $\sim 0.84$~Jy/beam, corresponding to a  brightness temperature of $>2 \times 10^5$~K. Hence, the bright feature at 16~\kms is undoubtedly a maser. Temporal variations in the \tn~line profile (Fig.~\ref{spec}) further support non-thermal processes in the excitation of \tn.

The \tr~emission is likely  maser in nature as well. 
For the strongest \tr~emission in the --4.8, --1.8, and 14.8 \kms channels, we infer a brightness temperature of $> 1200$~K (for a 0\pas2 beam), which is comparable to the value inferred for \tn~from maps with the same resolution. The estimated value is  well above the kinetic temperature observed even in the nearby hot core (150-300~K) (e.g., \citealt{Gen89}), but it could be plausible in principle in shocked gas, where the SiO molecule is believed to be produced and excited. 
In particular, the SiO molecule is produced in C-type shocks by the injection into the gas-phase of Si by grain sputtering and/or grain-grain collisions, followed by gas-phase reactions with O/O$_2$ \citep{Sch97a,Cas97}. At typical speeds of C-type shocks (10-50~\kms), the temperature of the post-shocked gas rises up to 1000~K and then progressively decreases to its pre-shock values. Multi-line SiO observations in protostellar jets give kinetic temperatures $<1000$~K (typical values in the range 100-500~K: \citealt{Nis07,Cab07}). Hence, if the emission were thermal,  the inferred brightness temperature  would be only marginally consistent with the temperatures that could be achieved via C-type shocks and/or with typically observed kinetic temperatures  in SiO protostellar jets. Based on similarities (e.g., compactness)  with all other masers lines (\te\ $v=1,2$ and \tn\ $v=0$), the \tr\ emission is more likely to be a maser.

For some of the weak \tn~and \tr~features we obtain brightness temperatures in the range 260-400~K, indicating that the isotopic $v=0$ emission imaged with the VLA is probably part thermal and part maser as for the \te~$v=0$ emission. The difference between the single-dish and interferometric spectral profiles indicates that the VLA in B-configuration does not have the brightness temperature sensitivity (RMS noise $T_b\sim$80~K) to detect the weaker and extended thermal component (Sect.~\ref{maps}).

Based on our single-dish spectra and interferometric images, we conclude that the $v=0$ emission from all species is a combination of thermal and maser components.

\begin{deluxetable*}{ccccccccc}
\tabletypesize{\scriptsize}
\tablewidth{0pc}
\tablecaption{Input parameters in the LVG model.}
\tablehead{
\colhead{T$_*^{(1)}$} & \colhead{R$_{\rm rad}^{(2)}$} &  \colhead{R$_{\rm mas}^{(3)}$}   & \colhead{W$^{(4)}$}  & \colhead{N$_{\rm H_2}$}  & \colhead{T$_k$} &  \colhead{$\chi [\rm{SiO}/\rm{H}_2]^{(5)}$} & \colhead{\te:\tn:\tr$^{(5)}$} & \colhead{$ d  {\rm V} / dr^{(6)}$}   \\
\colhead{(K)} & \colhead{(AU)} &   \colhead{(AU)}  & \colhead{}  & \colhead{(cm$^{-3}$)}   & \colhead{(K)}  &  & &  \colhead{(\kms~AU$^{-1}$)}   
}
\startdata
$10^4$ & 0.04-5 & 10-1000 & $10^{-1}-10^{-8}$ &$10^4-10^{12}$ & 600-2400 & $10^{-4}$ & 1:1/19.5:1/30.5 &0.5 \\
\enddata
\tablecomments{\\
$^{(1)}$ T$_*$ is the blackbody temperature of the radiative field.\\
$^{(2)}$  $R_{\rm rad}$ is the radius of the radiative source. \\
$^{(3)}$  $R_{\rm mas}$ is the distance to the maser region. \\
$^{(4)}$  W is the radiation geometric dilution factor. \\
$^{(5)}$ $\chi [\rm{SiO}/\rm{H}_2]$ and \te:\tn:\tr\ are the (solar) fractional abundances of SiO and isotopologues. \\
$^{(6)}$  $ d  {\rm V} / dr$ is the velocity gradient as measured by the velocity dispersion of $v=1$ emission. 
}
\label{inp}
\end{deluxetable*}
\subsection{Physical conditions from a radiative transfer modeling}
\label{lvg}

The ground-state isotopologue and the vibrationally excited state emissions appear to  arise from gas within $\sim 100$~AU of Source~I, but possibly at somewhat different locations within the volume.
We have undertaken preliminary radiative transfer calculations to assess whether  putative offsets might be due to excitation requirements for the different transitions.

We used a radiative transfer code based on the Large Velocity 
Gradient (LVG) model of \citet{Doe95},  that includes 
both  collisional and radiative pumping but does not treat effects 
of maser saturation or line overlap. We carried out model calculations
using 600 energy levels ($v=0-4;\ J=0-39$ for each isotopic species).
  The main input parameters for the LVG  model are reported in Table~\ref{inp}.  
The model background continuum assumes 
a blackbody function of  $T_* =10^4$~K. The radiation 
field is reduced by a geometric dilution factor, $W=1/4 (R^2_{\rm rad}/R^2_{\rm mas})$, where $R_{\rm rad}$ is the radius of the radiative source and $R_{\rm mas}$ is the distance to the maser region. The lower limit of $R_{\rm rad}$ corresponds to a B star  photospheric radius ($7 \times R_{\odot}$) \citep{Zom07} and the upper limit roughly approximates the radio continuum size \citep{Rei07}. 
We do not take into account emission from dust around Source~I as dust is likely to be sublimated at radii $<$ 100 AU, where high-density SiO in gas phase is present.
A velocity gradient $ d  {\rm V} / dr=0.5$~\kms AU$^{-1}$ is assumed based on the velocity dispersion  measured across the arms of the X in the $v=1$ emission (Fig.~\ref{ag575}).
We adopt the following relative (approximately solar) abundances: $[\rm{SiO}/\rm{H}_2] = 10^{-4}$, 
 [\te]/[\tn] $\sim 19.5$,  and [\te]/[\tr] $\sim 30.5$  \citep{Pen81}).
The Einstein coefficients are calculated from the data of \citet{Tip81}, while the collisional rate coefficients are taken from \citet{Bie83a,Bie83b}.

For simplicity, we assume that the velocity field is spherically-symmetric 
at any point in the medium, i.e. the logarithmic velocity gradient 
$ d (ln \ {\rm V}) / dr$ = 1.0. The photon escape probability is then given by 
$\beta = (1 - e^{-\tau_s})/\tau_s$,
where the Sobolev optical depth is
$\tau_s = (c^3/8 \pi \nu^3_{ul})A_{ul}[n_u - n_l g_u/g_l]/(d{\rm V}/dr)$, 
where $d{\rm V}/dr$ is the velocity gradient and $n_u$ and $n_l$ are populations
per sublevel. Calculated population inversions are represented in terms of 
the gain coefficient 
$\gamma = \lambda^2 \Delta n / 8 \pi t_s$,
where $\Delta n$ is the population inversion and  $t_s$  is the radiative lifetime of the transition.

The physical conditions required for the excitation of different 
maser species are summarized in Table~\ref{out}.
The \te~$v=0 \ J=1-0$ transition is inverted at relatively low density (N$_{\rm H_2}<10^7$~cm$^{-3}$), and it is never inverted in a strong, hot radiation field ($W < 0.001$), consistent with an inner radius of 200-1000~AU (Sect.~\ref{res}).  It occurs at a wide range of kinetic temperatures but it is optimized at $T_k < 1200$~K. 

The \te~$v=1,2 \ J=1-0$ transitions are never inverted for N$_{\rm H_2} < 10^7$~cm$^{-3}$ and they favor 
densities of $10^8-10^{10}$~cm$^{-3}$ ($v=1$) and   $10^9-10^{11}$~cm$^{-3}$ ($v=2$).  These results 
explain the spatial separation between the ground ($\gtrsim 100$~AU) and excited vibrational states 
($\lesssim 100$~AU). The $v=2$ transition is optimized at higher temperatures ($T_k \sim 2000$~K) 
and it is more strongly inverted in a strong, hot radiation field than the $v=1$  transition, which is apparently 
quenched for W$ > 0.01$. This is in agreement with the finding that $v=2$ masers tend to lie closer to the star than $v=1$ masers, although there is substantial spatial overlap (Sect.~\ref{res}).

The \tn~and \tr~$v=0 \ J=1-0$ transitions are inverted across a broad range of parameter space. 
Their emission is optimized for conditions similar to \te~$v=1,2$ (N$_{\rm H_2}=10^8-10^{11}$~cm$^{-3}$ 
and $T_k=1000-2000$~K), but can also occur more weakly under conditions conducive to \te~$v=0$ inversion (N$_{\rm H_2}<10^7$~cm$^{-3}$). 
These elements are in agreement with our finding that strong isotopic maser emission is excited within 100~AU from Source~I as is \te~$v=1,2$ maser emission, 
while extended weak isotopic emission may occur  at larger distances 
from Source~I in more extended structures, such as along the NE-SW outflow traced by the \te~$v=0$ emission 
(Sect.~\ref{maps}).


In conclusion, our LVG radiative pumping model explains the bulk
SiO maser emission characteristics of the five maser transitions observed
toward Orion BN/KL. However, although the large-scale  spatial and velocity distribution is similar for \te~$v=1,2$ and \tn/\tr~$v=0$ maser centroids, our data indicate differences in the small-scale spatial distributions and a certain degree of anticorrelation in intensity. In order to take into account these effects, future work would need to incorporate non-local radiative transfer and possibility of line overlap among the isotopologues in the modeling of the excitation of SiO masers \citep{Gon97,Her00}. 

%
\begin{deluxetable}{ccccccccc}
\tablewidth{0pc}
\tablecaption{Optimal LVG excitation conditions of SiO masers.}
\tablehead{
\colhead{Species} & \colhead{W} &  \colhead{N$_{\rm H_2}$}   & \colhead{T$_k$}   \\
\colhead{} & \colhead{} &  \colhead{(cm$^{-3}$)}   & \colhead{(K)}   
}
\startdata
\te~$v=0$ & \colhead{$0.0001$} & $<10^7$        & $<$ 1200\\
\te~$v=1$ & $<0.01$  & $10^8-10^{10}$ & $>$ 1500 \\
\te~$v=2$ & $0.01$   & $10^9-10^{11}$ & $>$ 2000 \\
\tn~$v=0$ & $0.01$   & $10^8-10^{11}$ &  $>$ 1500 \\
\tr~$v=0$ & $0.01$   & $10^8-10^{11}$ & $>$ 1500 \\
\enddata
\label{out}
\end{deluxetable}
%


\section{Summary}

We have used the VLA at 7~mm wavelength to image five rotational transitions  from three SiO isotopologues towards Orion BN/KL: 
$^{28}$SiO $v=0,1,2$ and $^{29}$SiO and $^{30}$SiO $v=0$ ($J=1-0$).
We have imaged for the first time the $^{29}$SiO and $^{30}$SiO $v=0$   maser emission with an angular resolution $\lesssim$0\pas2.

The main findings of this paper are listed below.
\begin{itemize}

\item
\te~$v=0$  maser emission traces a bipolar outflow driven by Source~I extended over $\sim 700$~AU along a NE-SW axis. 
\item
SiO isotopic emission arises from compact components with a spatial and velocity distribution  generally similar to the \te~vibrationally excited emission and traces material within 100 AU of Source~I.
\item
With resolution of 0\pas2, $\sim 70$\% of the isotopic emission appears to be resolved. The missing flux may arise in the outflow traced at larger radii by  \te~$v=0$  emission.
\item
LVG radiative transfer calculations predict similar temperatures and densities for the optimum maser excitation of both \tn/\tr~$v=0$ and \te~$v=1,2$ masers ($T=1000-2000$~K, $n_{\rm H_2} \sim 10^{10 \pm 1}$~cm$^{-3}$), significantly higher than  what is required for  \te~$v=0$ ($T\sim 600-1200$~K, $n_{\rm H_2} \sim 10^{6 \pm 1}$~cm$^{-3}$). However, weak emission from the isotopologues cannot be excluded at the lower temperatures and densities that support \te~$v=0$.
\item
Although the emission distributions for different isotopologues are generally similar (e.g., in radius), the peaks do not appear to overlap in detail (although 
 blending effects may contribute to the apparent offsets owing to the coarse interferometer beam). 

\end{itemize}

The apparent positional offset of  maser features from different transitions and/or isotopologues   suggests that non-local line overlaps should be included in the modeling of the excitation of SiO (isotopic) masers. Follow up modeling   will use a non-local spherical geometry radiative transfer code with line overlap.
Higher-resolution and higher sensitivity simultaneous observations of multiple maser lines/species with future facilities like EVLA and ALMA will be very important to make progress in the understanding of the gas dynamics around Source~I and  to better constrain physical conditions around Source~I on small-scales.

\acknowledgments{The data presented here were obtained under the VLA programs AG575 and AG776.
This material is based upon work supported by the National Science Foundation under Grant No. NSF AST 0507478.}

\end{document}